\documentclass[10pt,conference]{IEEEtran}
\usepackage[latin9]{inputenc}
\usepackage{array}
\usepackage{multirow}
\usepackage{graphicx}

\makeatletter

\providecommand{\tabularnewline}{\\}

\ifCLASSINFOpdf
\else
\fi
\usepackage{flushend}

\makeatother

\begin{document}

\title{Database Assisted Automatic Modulation Classification Using Sequential
Minimal Optimization}

\author{\IEEEauthorblockN{K. Pavan Kumar Reddy, K. Lakhan Shiva, K. Abhilash, Y. Yoganandam  } \IEEEauthorblockA{Department of Electrical and Electronics Engineering\\ Birla Institute of Technology and Science-Pilani, Hyderabad Campus\\ Hyderabad, Telangana State, India\\Email: \{pavan.kuntla,lakhanshiva,kadiri123.abhilash\}@gmail.com, yogi@hyderabad.bits-pilani.ac.in }}
\maketitle
\begin{abstract}
In this paper, we have proposed a novel algorithm for identifying
the modulation scheme of an unknown incoming signal in order to mitigate
the interference with primary user in Cognitive Radio systems, which
is facilitated by using Automatic Modulation Classification (AMC)
at the front end of Software Defined Radio (SDR). In this study, we
used computer simulations of analog and digital modulations belonging
to eleven classes. Spectral based features have been used as input
features for Sequential Minimal Optimization (SMO). These features
of primary users are stored in the database, then it matches the unknown
signal's features with those in the database. Built upon recently
proposed AMC, our new database approach inherits the benefits of SMO
based approach and makes it much more time efficient in classifying
an unknown signal, especially in the case of multiple modulation schemes
to overcome the issue of intense computations in constructing features.
In various applications, primary users own frequent wireless transmissions
having limited their feature size and save few computations. The SMO
based classification methodology proves to be over 99 \% accurate
for SNR of 15 dB and accuracy of classification is over 95 \% for
low SNRs around 5dB.\linebreak{}
\linebreak{}
 \textit{Keywords- Automatic modulation classification, cognitive
radio, Database, sequential minimal optimization, unknown signal detection,
primary user, spectral based features.}
\end{abstract}




\section{INTRODUCTION}


Cognitive Radio (CR) is a cutting edge technology on which rigorous
research and Development (R\&D) is occurring recently. CR enables
efficient usage of radio frequency spectrum. Empty spaces in both
licensed and unlicensed spectrum, commonly known as white spaces are
used by CR to achieve its set of objectives. The idea here is to sense
for a spectrum hole and use it until its rightful owner appears. The
CR then modifies its frequency to match another white space and possibly
another class of modulation scheme and coding. This phenomenon supposedly
may occur every now and then, imposing a heavy feedback to the receiver
every time the modulation scheme and the coding changes. One possible
solution to this problem is to use Digital Signal processing (DSP)
analysis called automatic modulation classification. These calculations
can determine the type of modulation effectively using particular
feature set. But most of the work in this direction is applied to
High Frequency (HF) noise and cannot be applied to Additive White
Gaussian Noise (AWGN) channels. {[}1{]} 

The goal of this study is to implement an automatic modulation classifier
in an AWGN environment. So, we designed the database assisted AMC,
analysed the need for fast classification and proposed a solution.
This paper proposes Sequential Minimal Optimization (SMO) to train
the Support vector machines (SVMs) to do classification by segregating
a mixture of both digital and analog modulation schemes into different
classes. Consequently, our contribution to Cognitive Radio research
lies in classifying an unknown signal in a time efficient manner,
through integration of database.

The paper is organized as follows: Section II explains sequential
minimal optimization and related work. Simulation of modulated signals,
feature extraction and modulation classification are presented in
Section III. Section IV is about proposed database assisted design
and algorithm. Results and discussions of proposed approach are presented
in Section V. Conclusion is drawn in VI.

\section{SEQUENTIAL MINIMAL OPTIMIZATION}

In recent years, there has been shown a lot of interest by researchers
in application of Support Vector Machines (SVMs), as they perform
well on a wide variety of problems related to pattern recognition.
But, there are some drawbacks in SVMs, due to which it is not widely
used by large research communities. First reason is that training
SVMs is slow, especially for large problems. Second explanation is
that they are complex, subtle, and can be implemented by an expert
engineer. 

SMO is widely used in training support vector machines (SVMs). The
amount of memory required for this algorithm is linear in the training
set size, due to which large training sets are handled with very less
complexity and also matrix computation is avoided. So, SMO spans somewhere
between linear and quadratic in the training set size for various
test problems, while the standard SVM algorithm spans somewhere between
linear and cubic in the training set size. SMO's computation time
is took over by SVM evaluation, hence SMO is the best algorithm for
linear SVMs. On real world sparse data sets, SMO can be approximately
more than 1000 times faster than the normal algorithm {[}2{]}.

An SMO model is a representation of the data sets as points in space.
They are mapped so that the data sets of the separate categories are
divided by a clear gap, which is as wide as possible. New data sets
are then mapped into that same space and predicted to belong to a
category based on which side of the gap they fall into. It trains
a support vector classifier using kernel functions such as Gaussian
and Polynomial kernels and takes minimal time in doing so {[}2{]}.

The \emph{$K(x,x_{i})$} are the support vectors used by SMO. $\alpha$
is the weight or lagrange multiplier. The output equation is given
as, 

\smallskip{}

$y=f(x)$; $f(x)=\sum_{k=1}^{m}\alpha_{k}.K(x,x_{k})+b$\hfill{}(1)

\smallskip{}

\subsection{Related works}

Azzouz and Nandi {[}3{]} have proposed spectral based features and
algorithms to classify both analog and digital modulated signals.
Popoola and Olst have extracted same set of features and proposed
ANN based classification system {[}4{]}. The key features proposed
by Nandi are later used in {[}5{]} {[}6{]} and many more. But in practicality,
accuracy of these features depends upon noise, carrier offset error,
symbol rate estimation error that affect the probability of identification.
Arulampalam et al., {[}13{]} have used only digital modulation schemes
using ANN and features proposed by Nandi and Azzouz, validated over
non-weak segment of the incepted signal and also used maximum power
spectral density of normalized instantaneous frequency of intercepted
signal. When we compare our work with the study, we have obtained
better results by using SMO Algorithm {[}8{]} {[}15{]} for classificaion
of both analog and digital modulations at low SNRs. Our contribution
is in building upon this existing work on AMC, via a database approach
and classifying the unknown signal in a time efficient manner.

\section{STUDY METHODOLOGY}

The study flow involved three steps. The first step was extraction
of the statistical features keys used as inputs to the developed classifier.
The second step was development of automatic modulation classifier
system. Third step was to integrate the system with the database.
Satisfactory accuracy was achieved in the final step.

\subsection{Simulation of modulated signals}

Signals are generated for eleven modulation schemes, which includes
both analog and digital modulations. Modulated signals are generated
using the formulas. For example, the AM modulated signal is generated
using the formula,

\smallskip{}

\smallskip{}

$Y_{am}(t)=[1+K_{a}.F_{t}](\cos(2\pi f_{c}t))$ \hfill{}(2)

\smallskip{}

\smallskip{}

The FM modulated signal is generated using the formula,

\smallskip{}

\smallskip{}

$Y_{fm}(t)=\cos(2\pi f_{c}t+k_{f}\sin(2\pi f_{m}t))$ \hfill{}(3) 

\smallskip{}
\smallskip{}

Similarly DSB, SSB, ASK, FSK and PSK modulated signals in an AWGN
environment were generated in MATLAB using standard formulas.

\subsection{Feature extraction}

Feature extraction is a prerequisite for recognition. The motive here
is to identify a pattern by using minimum number of features or attributes,
which are important for determining the pattern class identification.
The employed classification technique was derived from the amplitude,
phase and frequency instantaneous values of signal simulations. These
values are obtained from the discrete time analytic signal. Hilbert
transform is used for calculating the instantaneous values of a signal.
The key features used in this paper are spectral based features and
had earlier been used by Nandi {[}3{]} {[}7{]}.

(i) $\gamma_{max}$ : This feature corresponds to maximum value of
the PSD (Power spectral density) . It measures the variance in signal
amplitude. 

\smallskip{}

\smallskip{}

$\gamma_{max}=\frac{max|DFT(a_{cn}[n]|^{2}}{N}$ \hfill{}(4)

\smallskip{}

\smallskip{}

i.e. taking a discrete Fourier transform of the signal samples. Here
$a_{cn}(n)$ is centered instantaneous amplitude at time, $t=\frac{n}{f_{s}}$($n=1,2,3...N$)
and a sampling frequency, $f_{s}$. $a_{cn}(n)$ is defined as 

\smallskip{}

\smallskip{}

$a_{cn}(n)=a_{n}[n]-1$ , $a_{n}(n)=\frac{a[n]}{m_{a}}$ \hfill{}(5)

\smallskip{}
\smallskip{}

Here $m_{a}$ is the mean value of samples. 

\smallskip{}

\smallskip{}

$m_{a}=\frac{1}{N}\sum_{n=1}^{N}a[n]$ \hfill{}(6)

\smallskip{}

\smallskip{}

All centered values refer to normalized instantaneous values.

\smallskip{}
\smallskip{}

(ii) $\sigma_{dp}$ : This feature measures the variance in direct
instantaneous phase. It discriminated 2PSK from other modulation schemes.
This is given as $\sigma_{dp}$

\smallskip{}

\smallskip{}
$\sigma_{dp}=\sqrt{\frac{1}{N_{c}}\left(\sum_{A_{n}>A_{t}}\Phi_{NL}^{2}[n]\right)-\left(\frac{1}{N_{c}}\sum_{A_{n}>A_{t}}\Phi_{NL}[n]\right)^{2}}$

\hfill{}(7)

\smallskip{}

\smallskip{}

(iii) $\sigma_{ap}$ : This feature corresponds to the standard deviation
of the centered non-linear component of the instantaneous phase. Here
we take absolute values. This is given as 

\smallskip{}

\smallskip{}
 $\sigma_{ap}=\sqrt{\frac{1}{N_{c}}\left(\sum_{A_{n}>A_{t}}\Phi_{NL}^{2}[n]\right)-\left(\frac{1}{N_{c}}\sum_{A_{n}>A_{t}}|\Phi_{NL}[n]|\right)^{2}}$

\hfill{}(8)

This is very useful in finding PSK modulation order.

\smallskip{}

\smallskip{}

(iv) $P$ : This feature is the calculation of spectrum symmetry around
the carrier frequency. This quantity is a decision factor to classify
amplitude-based modulations with different properties in the frequency
domain.

\smallskip{}

\smallskip{}

$P=\frac{P_{l}-P_{u}}{P_{l}+P_{u}}$\hfill{}(9)

\smallskip{}

\smallskip{}

$P_{l}$is the spectral power of sideband (lower) and $P_{u}$is the
spectral power of the other sideband (upper). 

\smallskip{}

\smallskip{}

$P_{l}=\sum_{n=1}^{f_{cn}}|F(n)|^{2}$, $P_{u}=\sum_{n=1}^{f_{cn}}|F\left[n+f_{cn}+1\right]|^{2}$
\hfill{}(10)

\smallskip{}

\smallskip{}

Here $F(n)$ is the Fourier transform of the received signal. ($f_{cn}+1$)
is the sample number for carrier frequency $f_{c}$ . $f_{cn}$ is
defined as 

\smallskip{}

\smallskip{}

$f_{cn}=\frac{f_{c}N}{f_{s}}-1$ \hfill{}(11)

\smallskip{}

\smallskip{}

(v) $\sigma_{aa}$ : This feature corresponds to the standard deviation
of the centered instantaneous amplitude of signal samples. $N$ is
the total number of signal samples. We take absolute and normalized
values. Though this feature is similar to $\gamma_{max}$, it has
the ability to discriminate 2ASK from other modulation schemes.

\smallskip{}

\smallskip{}

$\sigma_{aa}=\sqrt{\frac{1}{N}\left(\sum_{n=1}^{N}A_{cn}^{2}(n)\right)-\left(\frac{1}{N}\sum_{n=1}^{N}|A_{cn}(n)|\right)^{2}}$\hfill{}(12)

\smallskip{}
\smallskip{}

(vi) $\sigma_{af}$ : This feature corresponds to standard deviation
of the centered instantaneous frequency. We take absolute and normalized
values. It has the ability to discriminate 2ASK from 4ASK. $N_{c}$
is the number of samples that meet the condition: $A_{n}[n]>A_{t}$.
$A_{t}$ is the threshold value.

\smallskip{}

\smallskip{}

$\sigma_{af}=$$\sqrt{\frac{1}{N_{c}}\left(\sum_{A_{n}[n]>A_{t}}f_{N}^{2}(n)\right)-\left(\frac{1}{N_{c}}\sum_{A_{n}[n]>A_{t}}|f(n)|\right)^{2}}$

\hfill{}(13)

\smallskip{}

\smallskip{}

(vii) $\sigma_{a}$: This feature corresponds to the standard deviation
of the centered frequency. We take normalized values here. It has
the ability to differentiate 2ASK and 4ASK. 

\smallskip{}

\smallskip{}

$\sigma_{a}=\sqrt{\frac{1}{N_{c}}\left(\sum_{A_{n}[n]>A_{t}}a_{cn}^{2}[n]\right)-\left(\frac{1}{N_{c}}\sum_{A_{n}[n]>A_{t}}a_{cn}[n]\right)^{2}}$

\hfill{}(14)

\smallskip{}

\smallskip{}

(viii) $\mu_{42}^{a}$: This feature is the kurtosis of normalized
and centered instantaneous amplitude. It differentiates AM signal
from the rest. 

\smallskip{}

\smallskip{}

$\mu_{42}^{a}=\frac{E\{A_{cn}^{4}[n]\}}{\{E\{A_{cn}^{2}[n]\}\}^{2}}$\hfill{}(15)

\smallskip{}

\smallskip{}

(ix) $\mu_{42}^{f}$: This feature is the kurtosis of normalized and
centered instantaneous frequency. It differentiates FM signal from
the rest. 

\smallskip{}

\smallskip{}

$\mu_{42}^{f}=\frac{E\{f_{N}^{4}[n]\}}{\{E\{f_{N}^{2}[n]\}\}^{2}}$\hfill{}(16)

\smallskip{}

\smallskip{}

\subsection{Modulation classification}

Modulation can be defined as varying amplitude, frequency and phase
of a carrier wave with frequency $f_{c}$ with respect to modulating
signal with frequency $f_{m}${[}6{]}.

While training the SVM, nominal attributes in the arff format are
converted into two class (binary) ones. Coefficients in the output
are based on the normalized data. The SMO algorithm will use pair
wise classification considering the fact that the problem here is
a multi-class one. We choose to make use of Poly Kernel i.e. polynomial
kernels. Here since we have 11 classes the 11 binary SMO models have
been output with one hyperplane each to support each of the possible
pair of class values. In such Kernel based methods, similarity must
be a kernel function satisfying some mathematical properties {[}11{]}
{[}12{]}.

\section{DESIGN AND ALGORITHM}

The steps in design of the database assisted AMC are presented in
Fig. 1. 

\begin{figure}

\includegraphics[width=3.5in,height=3.5in]{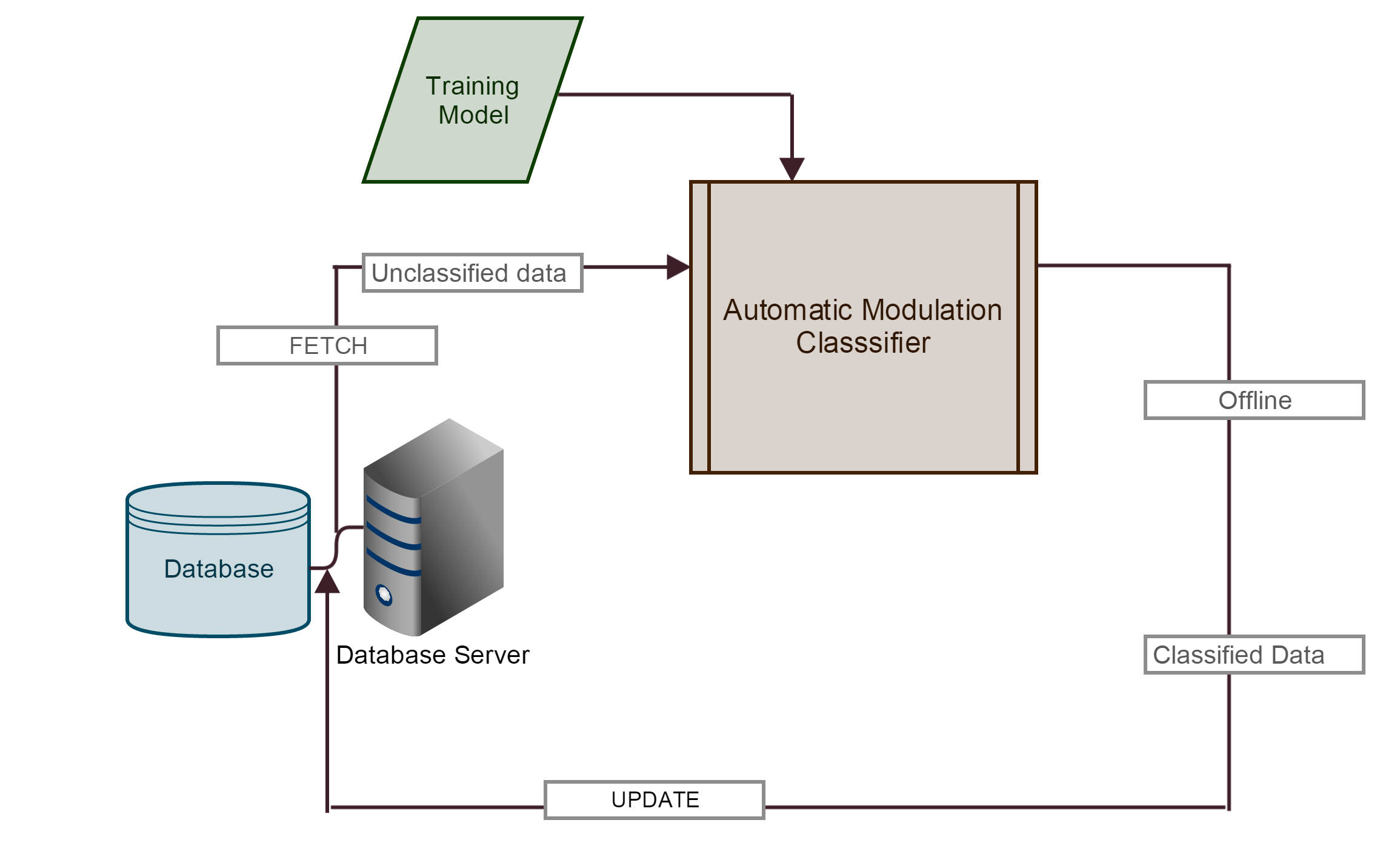}

\hfill{} Fig. 1. Database assisted Automatic modulation classifier\hfill{} 
\end{figure}

\smallskip{}

\smallskip{}

The steps for implementing the Database assisted AMC system are divided
into two phases. 

1) Offline Phase: 

\begin{itemize}

\item[$\bullet$]Firstly we had realized computer simulated signals
of different modulation schemes using MATLAB codes reproducing the
work of A.K. Nandi and E.E. Azzouz {[}3{]}.

\item[$\bullet$]Then, stored the simulated and labeled signal data
into the database server. 

\item[$\bullet$]Thereafter we have generated reference sample data,
to train the SMO model offline. 

\item[$\bullet$]Further saved the trained model into the file system.

\item[$\bullet$]Finally evaluated by giving an unlabeled data as
input to the database server (this will be used as test data that
will be classified by AMC).

\end{itemize}

2) Classification Phase: 

\begin{itemize}

\item[$\bullet$]Secondly we have connected to the database using
JDBC MySQL connector/J. 

\item[$\bullet$]Accordingly fetched all the unlabeled data into the
program. 

\item[$\bullet$]Next we attached arff (attribute relation file header)
header format to this data in order to transform it into an arff test
file. 

\item[$\bullet$]Then read the testing file into the program and loaded
the trained SMO model. 

\item[$\bullet$]Afterwards we have applied the model's configuration
on the test file. With this step, modulation classification is completed.
In the end, updated the database server with the classified signal
data. 

\end{itemize}

\smallskip{}

\smallskip{}

\section{RESULTS AND DISCUSSIONS}

\begin{figure}
\includegraphics[clip,width=3.2in,height=1.7in]{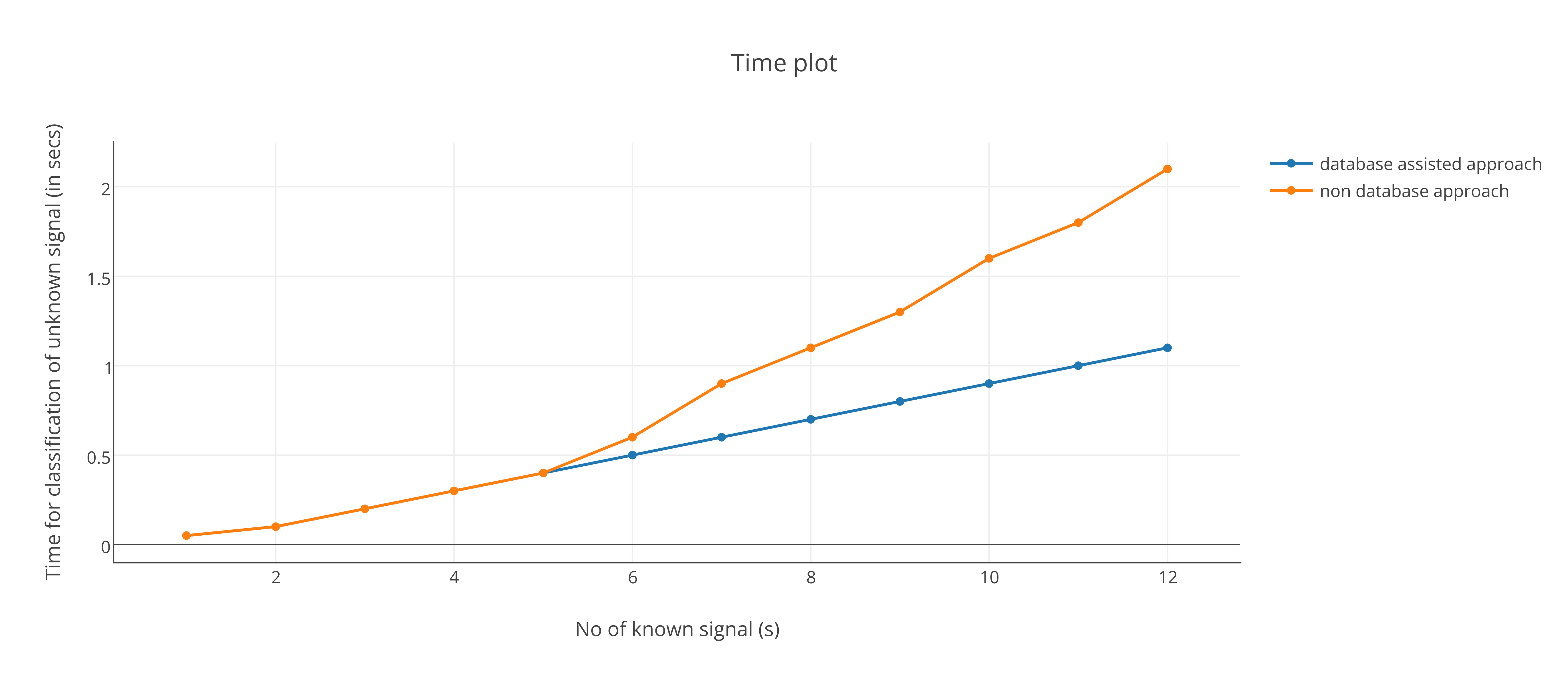}

\hfill{} Fig. 2. Time for classification of unknown signal vs. number
of known signals.\hfill{} 
\end{figure}

\smallskip{}

Here we have input test data (unknown signal) of 25dB SNR and applied
an SMO trained model over it. The model uses some hyper planes to
divide the space to form 11 regions, each representing one modulation
scheme. This has been explained in section II of this paper.

In database assisted approach, when an unknown signal's features are
among the training data stored in database, an existing record in
the database is selected and its modulation scheme is returned. We
have first trained the SMO classifier and stored the trained model
in WEKA {[}9{]} and database. Finally, we are classifying unknown
signal with an assumption that it should belong to one of the known
modulation schemes mentioned in section III of this paper.

In our database assisted approach, we record the features of known
signals in the database system. Then, AMC monitors each incoming signal\textquoteright s
transmitting sequence. For example, let us suppose that initially
only one known signal, such as AM is present in an environment and
detect the unknown signal. Then, in the next input sequence consider
only two known signals such as AM, FM and compare the unknown signal\textquoteright s
feature vectors with those in the database. In various applications,
known signals have routine wireless transmissions, hence they own
a limited number of feature vectors, meaning that the subsequent database
is stable and limited in size. Whenever an unknown signal's feature
vector does not have a match entity in the database, our approach
will detect it's action as unusual by taking the stability of known
signals into account and identifies it as a malicious signal, hence
save some computations compared to the previous approaches in the
literature. This way we realized that our proposed idea on database
assisted approach for AMC is a novel work. So, we ran computer simulations
and in future would like to implement this idea on a system model
similar to cognitive radio scenario of known signals. Known signals
are analogous to primary users' signals and an unknown signal referred
as primary user is to be detected so that interference can be avoided
with neighboring primary users. Therefore, we assumed all basic modulation
schemes possible as the only modulation types used by authentic primary
users' signals, referred as known signals. Here our research contribution
is to evaluate the performance of database approach over non database
in terms of efficiency, thereby solving the problem of severe computations
in constructing features. 

Our limitations are as follows: 

\begin{enumerate}

\item At any instance of time, only one signal is transmitted by
user's either primary, unknown or secondary.

\item Secondary user's transmit at very low powers not interfering
with primary and unknown users.

\item Secondary users, unknown users and primary users are utilizing
the area within the same frequency band.

\item Secondary user's modulation scheme is different from primary
and unknown users.

\item We cannot detect intelligent unknown user's who pretend to
be primary users.

\item We have not considered other modulation schemes for transmission
of unknown signal, instead used only those modulation schemes mentioned
in section III of this paper.

\end{enumerate}

In non database approach, we have assumed that the test signal's feature
is among the training data. To match this test signal to a record
in the arff format, we need to search for the record. But running
a search query is not preferred in flat files (File that doesn't have
any internal hierarchy). The reason is that database is a structured
way of storing data, whereas flat file is unstructured way of storing
data.

It is very clear from Fig. 2 that for both database and non database
methods the time taken for classifying an unknown signal is surely
dependent on number of known signals. When there are more known signals
in the system, it consumes more time for conclusion. However, it is
observed that with a larger number of known signals, the classification
time increases more drastically in the case of non database approach.
The database assisted approach takes linear growth, because time taken
for classification is dominated only by searching time of database.
Whereas, non database approach follows approximately exponential rise,
due to increase in classification time with number of known signals.

At low SNRs like 5 dB, as signals are transmitted with low power,
it is a challenge to classify 11 different modulation schemes, as
sometimes there is confusion to classify the similar modulation schemes.
Generally, we get accuracy of 80-90\% at low SNRs by recent methods.
But by these features we are getting around 95\%.

\subsection{Probability of correct detection }

In multi-class prediction, the result on a test set is often displayed
as a two dimensional confusion matrix with a row and column for each
class. It is defined by this formula, $P=(TP+TN)/(TP+TN+FP+FN)$.
Here $TP$ means True Positives, $TN$ means True negatives $FP$
means False positive and $FN$ means False negative. The error rate
is one minus this {[}10{]}.

\pagebreak{}

\hfill{}TABLE I. CORRECT DETECTION RATE\hfill{}

\smallskip{}

\smallskip{}

\hfill{}%
\begin{tabular*}{2.7in}{@{\extracolsep{\fill}}|>{\raggedright}p{1.8cm}|>{\centering}p{1.8cm}|>{\centering}p{1.8cm}|}
\hline 
\multirow{3}{1.8cm}{Modulation Scheme} & \multicolumn{2}{c|}{Accuracy}\tabularnewline
\cline{2-3} 
 & \multicolumn{2}{c|}{SNR = 15 dB}\tabularnewline
\cline{2-3} 
 & Present Work & Popoola Work {[}4{]} \tabularnewline
\hline 
2ASK & 100 & 99.99\tabularnewline
\hline 
4ASK & 100 & 99.98\tabularnewline
\hline 
2FSK & 100 & 99.92\tabularnewline
\hline 
2PSK & 100 & 99.75\tabularnewline
\hline 
4PSK & 100 & 99.98\tabularnewline
\hline 
AM & 100 & 99.95\tabularnewline
\hline 
DSB & 99.97 & 99.98\tabularnewline
\hline 
FM & 100 & 99.91\tabularnewline
\hline 
LSB & 100 & 99.97\tabularnewline
\hline 
USB & 99.97 & -\tabularnewline
\hline 
\end{tabular*}\hfill{}

\smallskip{}

\smallskip{}

\smallskip{}

\smallskip{}

\hfill{}TABLE II. CONFUSION MATRIX\hfill{}

\smallskip{}

\smallskip{}

\smallskip{}

\hfill{}%
\begin{tabular*}{3.5in}{@{\extracolsep{\fill}}|c|c|c|c|c|c|c|}
\hline 
 & \multicolumn{6}{c|}{Confusion matrix for SNR = 15 dB}\tabularnewline
\hline 
\hline 
 & AM & DSB & LSB & USB & FM & 2ASK\tabularnewline
\hline 
AM & 1 & 0 & 0 & 0 & 0 & 0\tabularnewline
\hline 
DSB & 0 & 1 & 0 & 0 & 0 & 0\tabularnewline
\hline 
LSB & 0 & 0 & 1 & 0 & 0 & 0\tabularnewline
\hline 
USB & 0 & 0 & 0.01 & 0.99 & 0 & 0\tabularnewline
\hline 
FM & 0 & 0 & 0 & 0 & 1 & 0\tabularnewline
\hline 
2ASK & 0 & 0 & 0 & 0 & 0 & 1\tabularnewline
\hline 
\end{tabular*}\hfill{}

\smallskip{}

\smallskip{}

\smallskip{}

\smallskip{}

\hfill{}TABLE III. CONFUSION MATRIX\hfill{}

\smallskip{}

\smallskip{}

\hfill{}%
\begin{tabular*}{3.5in}{@{\extracolsep{\fill}}|c|c|c|l|c|c|}
\hline 
 & \multicolumn{5}{c|}{Confusion matrix for SNR = 15 dB}\tabularnewline
\hline 
\hline 
 & 4ASK & 2FSK & 4FSK & 2PSK & 4PSK\tabularnewline
\hline 
4ASK & 1 & 0 & 0 & 0 & 0\tabularnewline
\hline 
2FSK & 0 & 1 & 0 & 0 & 0\tabularnewline
\hline 
4FSK & 0 & 0 & 1 & 0 & 0\tabularnewline
\hline 
2PSK & 0 & 0 & 0 & 1 & 0\tabularnewline
\hline 
4PSK & 0 & 0 & 0 & 0 & 1\tabularnewline
\hline 
\end{tabular*}\hfill{}

\smallskip{}

\smallskip{}

Table I illustrates the improvement of present work with reference
work by Popoola's {[}4{]} {[}16{]} feed forward neural networks {[}14{]}.
Table II and III illustrate the confusion matrix for different modulation
schemes at an SNR = 15 dB. 

\smallskip{}

\smallskip{}

\section{CONCLUSION}

In this paper, we have developed and implemented an AMC with database
acting as an abettor in automating the whole process of classification.
The database assisted approach gave us better results than the non
database approach in terms of time taken. We have applied SMO model
configuration on both labeled and unlabeled data. Firstly, SMO based
classification alone gave better performance when trained and evaluated
with labeled signals. The results proved to be over 98\% accurate
for 15 dB data. The future work of this paper includes validation
of our approach through hardware implementation to show its feasibility
in real life conditions.

In future, we will also consider same modulation type during classification.
Till now we are assuming that the unknown signal's modulation should
belong to any one of the 11 modulation schemes generated in simulations.
But, in real life we need to detect an unknown signal, which may be
different from modulations stored in database. Then, we consider such
an unknown signal as malicious. Now, as malicious users (unknown users)
are intelligent, they pretend modulations of primary users. So, in
this case, if both primary and unknown users use same modulation type,
and no match is found in database. Then, during the detection phase
of unknown signals, one can employ a new method, which will detect
these type of unknown signals.






\bibliographystyle{IEEEtran}
\bibliography{C:/Users/DELL/Downloads/citations}

\nocite{zhang2011introduction,plattsequential,nandi1997modulation,popoola2011automatic,nandi1998algorithms,azzouz1996procedure,zhu2014automatic,chaithanya2010blind,quinlan2014c4,ware2000weka,witten2005data,tabatabaei2010svm,arulampalam1999classification,guldemir2007online,park2008automatic,popoola2011novel}

\end{document}